\documentclass{PoS}

\title{Calculating the running coupling in strong electroweak models}

\ShortTitle{Calculating the running coupling in strong electroweak models}

\author{Zolt\'{a}n Fodor \\
Department of Physics, University of Wuppertal, Gauss
  Strasse 20, D-42119, Germany \\
  and \\
  Institute for Theoretical Physics, Eotvos University Budapest,
  Pazmany P. 1, H-1117, Hungary \\ 
E-mail: \email{fodor@bodri.elte.hu}}

\author{\speaker{Kieran Holland} \\
Department of Physics, University of the Pacific\\
  3601 Pacific Ave, Stockton CA 95211, USA \\
E-mail: \email{kholland@pacific.edu}}

\author{Julius Kuti \\
  Department of Physics 0319, University of California, San Diego \\
  9500 Gilman Drive, La Jolla CA 92093, USA \\
E-mail: \email{jkuti@ucsd.edu}}

\author{D\'{a}niel N\'{o}gr\'{a}di \\
  Department of Physics 0319, University of California, San Diego \\
  9500 Gilman Drive, La Jolla CA 92093, USA \\
E-mail: \email{nogradi@bodri.elte.hu}}

\author{Chris Schroeder \\
  Department of Physics 0319, University of California, San Diego \\
  9500 Gilman Drive, La Jolla CA 92093, USA \\
E-mail: \email{crs@physics.ucsd.edu}}

\abstract{One possibility for Beyond Standard Model physics is a new
  strongly-interacting gauge theory. One way to determine if a
  non-abelian gauge theory is QCD-like or conformal is to measure the
  running of the renormalized gauge coupling. We define the
  renormalized coupling from Wilson loop ratios, and measure these
  ratios via lattice simulations. We test this method in $SU(3)$ pure
  gauge theory and show some first results for simulations with
  dynamical fermions in the fundamental representation.}

\FullConference{The XXVII International Symposium on Lattice Field Theory\\
		 July 26-31, 2009\\
		 Peking University, Beijing, China}

\newcommand{\be}{\begin{equation}}
\newcommand{\ee}{\end{equation}}
\newcommand{\bea}{\begin{eqnarray}}
\newcommand{\eea}{\end{eqnarray}}
\def\nn{\nonumber}

\begin{document}

\section{Running coupling}

There is great interest in the possibility that Beyond Standard Model
physics might take the form of new strongly coupled gauge theories
\cite{Weinberg:1979bn}--\cite{Appelquist:2003hn}, one example being
technicolor. For model building, it is necessary to distinguish
conformal theories from those with a mass gap like QCD. There have
been a number of lattice studies of this issue
\cite{Appelquist:2007hu}--\cite{Bilgici:2009kh}. One signal is the
running coupling, which has an infrared fixed point in a conformal
theory. We developed our method for the running coupling and its beta
function from the definition of the continuum renormalized coupling
constant using the second derivative of $R\times T$ Wilson loops
\cite{Fodor:2009wk}. This coupling runs with the size of the Wilson
loop in infinite volume~\cite{Campostrini:1994fw}. We generalized this
definition to a finite volume $L^4$ keeping $R/L$ fixed and run the
coupling with $L$ as in the Schrodinger functional method. A similar
method was developed independently in~\cite{Bilgici:2009kh}.  

Consider Wilson loops $W(R,T,L)$, where $R$ and $T$ are the space-like
and time-like extents of the loop, and the lattice volume is
$L^4$ (all dimensionful quantities are expressed in units of the
lattice spacing $a$). A renormalized coupling can be defined by
\be g^2(R/L,L) = 
  - \frac{R^2}{k(R/L)} \frac{\partial^2}{\partial
  R \partial T} \ln \langle W(R,T,L) \rangle \left. \right|_{T=R},
\ee
where for convenience the definition will be restricted to Wilson loops with
$T=R$, and  $\langle ... \rangle$ is the expectation value of some
quantity over the full path integral. This definition can be motivated
by both renormalized and bare perturbation theory, where the leading
term is the tree-level coupling. The renormalization scheme is defined
by holding $R/L$ to some fixed value. The quantity $k(R/L)$ can be
calculated from Wilson loop expectation values using
perturbation theory. This is done numerically on finite lattices,
hence $k$ contains lattice artifacts which vanish as $L \rightarrow
\infty$. The role of lattice simulations is to measure
the expectation values non-perturbatively. On the
lattice, derivatives are replaced by finite differences, so 
the renormalized coupling is defined to be 
\bea
 && g^2((R+1/2)/L,L) = 
\frac{1}{k(R/L)} (R+1/2)^2 \chi(R+1/2,L)\, , \nn \\
&& \chi(R+1/2,L) = 
-\ln \left[ \frac{W(R+1,T+1,L)W(R,T,L)}{W(R+1,T,L)W(R,T+1,L)} \right]
\left. \right|_{T=R},
\label{eq:g}
\eea
where $\chi$ is the Creutz ratio~\cite{Creutz:1980zw}, 
and the renormalization scheme is defined by holding the value of
$r=(R+1/2)/L$ fixed. 

With this definition, the renormalized coupling $g^2$ is a function of
the lattice size $L$ and the fixed value of $r$. The continuum limit
corresponds to $L \rightarrow \infty$, where the physical length
scale $L_{\rm phys}$ is held fixed while the lattice spacing $a
\rightarrow 0$. The coupling is non-perturbatively defined, as the
expectation values are calculated via lattice simulations, which
integrate over the full phase space of the theory. One starts the RG
flow from some reference physical point $L_{{\rm phys},0}$ which is set by the
choice e.g.~$g^2(r,L_{{\rm phys},0})=0.8$. In a QCD-like theory, $g^2$ increases
with increasing $L_{\rm phys}$ flowing in the infrared direction. In a conformal
theory, $g^2$ flows towards some non-trivial infrared fixed point
$g^{*2}$ as $L_{\rm phys}$ increases, whereas in a trivial theory, $g^2$
decreases with $L_{\rm phys}$. 

One way to measure the running of the renormalized coupling in the
continuum limit is via step-scaling. The bare lattice coupling is
defined in the usual way $\beta = 6/g_0^2$ as it appears in the
lattice action. On a sequence of lattice sizes
$L_1, L_2, ..., L_n$, the bare coupling is tuned on each lattice so that
exactly the same value $g^2(r,L_i,\beta_i)=g^2(r,L_{\rm phys})$ is measured 
via simulations. Next a new set of simulations is performed, on a sequence of
lattice sizes $2L_1, 2L_2, ..., 2L_n$, using the corresponding tuned
couplings $\beta_1, \beta_2, ..., \beta_n$. From the simulations, one
measures $g^2(r,2L_i, \beta_i)$, which vary with the bare coupling
i.e.~the lattice spacing. These data are extrapolated to the
continuum as a function of $1/L_i^2$. This gives one blocking step
$g^2(r,L_{\rm phys}) \rightarrow g^2(r,2L_{\rm phys})$ in the
continuum RG flow. The whole procedure is then iterated.  The chain of
measurements gives the flow $g^2(r,L_{\rm phys}) \rightarrow
g^2(r,2L_{\rm phys}) \rightarrow g^2(r,4L_{\rm phys}) \rightarrow
g^2(r,8L_{\rm phys}) \rightarrow ...$, as far as is feasible.  One is
free to choose a different blocking factor, say $L_{\rm phys}
\rightarrow (3/2)L_{\rm phys}$, in which case more blocking steps are
required to cover the same energy range. 

\begin{figure}
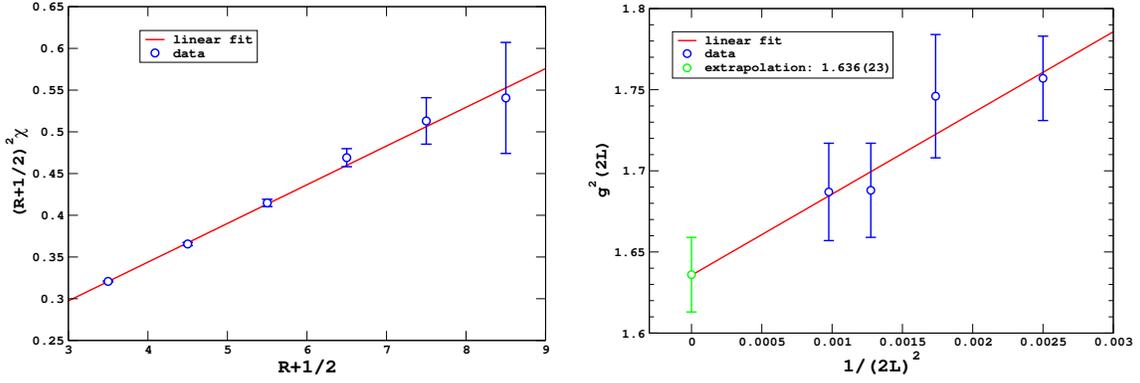

\begin{center}
\begin{tabular}{c c}
\vspace{0.5truecm}
\hspace{-0.2cm}
\includegraphics[width=0.48\textwidth,angle=0]{creutz.l28.b6.99.grace.v3.eps}
\hspace{0.3cm}
\includegraphics[width=0.48\textwidth,angle=0]{gsq2L.gsqL.1.44.grace.eps}
\end{tabular}
\end{center}
\vspace{-0.6truecm}
\caption{(Left) The rescaled Creutz ratio on a $28^4$ lattice at
  $\beta=6.99$. We interpolate the data linearly to
  $r=(R+1/2)/L=0.25$, giving a chi squared per degree of freedom
  2.0/4. (Right) The measured coupling $g^2(2L_i,\beta_i)$ for
  $2L_i=20, 24, 28$ and 32, where $\beta_i$ is tuned such that
  $g^2(L_i,\beta_i)=1.44$. A linear continuum extrapolation gives
  $g^2(2L_{\rm phys})=1.636(23)$, with $\chi^2$/dof = 0.57/2. }
\label{fig1}
\end{figure}

\section{ $SU(3)$ pure gauge theory}

As our first test of this method, we study $SU(3)$ pure gauge theory
in four dimensions. We simulate using the standard Wilson lattice
gauge action, with a mixture of five over-relaxation updates for every
heatbath update. We define the renormalization scheme with the fixed
value $r=0.25$, for brevity we omit the label $r$ in the renormalized
coupling. In this $R/L$ range, the Creutz ratio can be 
accurately measured, and the geometric factor $k$ converges quickly to
its continuum-limit value. In the pure gauge theory test, we actually
use the finite $L$ values of $k$, as this may remove some of the
cutoff dependence of the renormalized coupling. For the RG flow we
choose the blocking step $L \rightarrow 2L$. We simulate on small
lattices of size $L=10, 12, 14, 16, 18, 20$ and 22, and the
corresponding doubled lattices $2L=20, 24, 28, 32, 36, 40$ and
44. Wilson loop measurements are separated by 1000 sweeps, which we
find is sufficient to generate statistically independent configurations. 

To tune $\beta_i$ for each small lattice size $L_i$, we typically run 
separate simulations at 5 -- 10 different $\beta$ values in the
relevant range of renormalized coupling $g^2$. Each of these runs
contains 300 -- 500 measurements i.e.~up to $5 \times 10^5$
sweeps. Simulating on the doubled lattices at the tuned $\beta_i$
values, we generate between 200 and 1000 measurements each, typically
more than 500. The signal of the Creutz ratio disappears into the
noise as the size  of the Wilson loop increases. One way to suppress
the noise in measurements is to gauge fix the configurations to
Coulomb gauge, and replace the thin-link Wilson loop with the
correlator of the products of the time-like gauge links
\cite{Bazavov:2009bb,Bernard:2000gd}. Note that gauge fixing is not
implemented in the actual Monte Carlo updating algorithm. An
alternative method to suppress noise is to smear the gauge links and
measure the fat-link Wilson loop operator. In the pure gauge theory
test, we use the gauge-fixing method, in the dynamical fermion
simulations we describe later, we use the smearing method. These
improvement methods do not correspond to calculating the original
thin-link Wilson loop operator. 

We show in Fig.~\ref{fig1}~(left) a typical result for the rescaled
Creutz ratio $(R+1/2)^2 \chi$. The doubled lattice is $28^4$ and the
bare coupling $\beta=6.99$ is tuned from simulations on $14^4$
volumes. Errorbars are calculated using the jackknife method. The
renormalized coupling is defined at the point $r=(R+1/2)/L=0.25$,
corresponding to $(R+1/2)=7$ at this lattice size. We interpolate the
data linearly to this point, obtaining $\chi^2$/dof$=2.0/4$. The data at
different $R$ are highly correlated, being measured on the same gauge
configurations. To calculate an error, we bin the gauge
configurations, and analyze and interpolate separately each bin,
giving a distribution of interpolated values.
An example of the step-scaling method is shown in
Fig.~\ref{fig1}~(right). The bare couplings are tuned such that
$g^2(L_i,\beta_i)=1.44$ for $L_i=10, 12, 14$ and 16, the figure shows
the data $g^2(2L_i, \beta_i)$ for $2L_i=20, 24, 28$ and 32. The
leading lattice artifacts in the Creutz ratio are expected to be of
order ${\cal O}(a^2)$. This corresponds to ${\cal O}(1/L^2)$, since
the physical lattice size is $La$. Extrapolating linearly in
$1/(2L)^2$ gives $g^2(2L_{\rm phys})=1.636(23)$ and $\chi^2$/dof = 0.57/2. For a
systematic error, we omit one data point at a time and repeat the
extrapolation. Combined in quadrature with the statistical error, our
continuum result is $g^2(2L_{\rm phys})=1.636(25).$ 

\begin{figure}
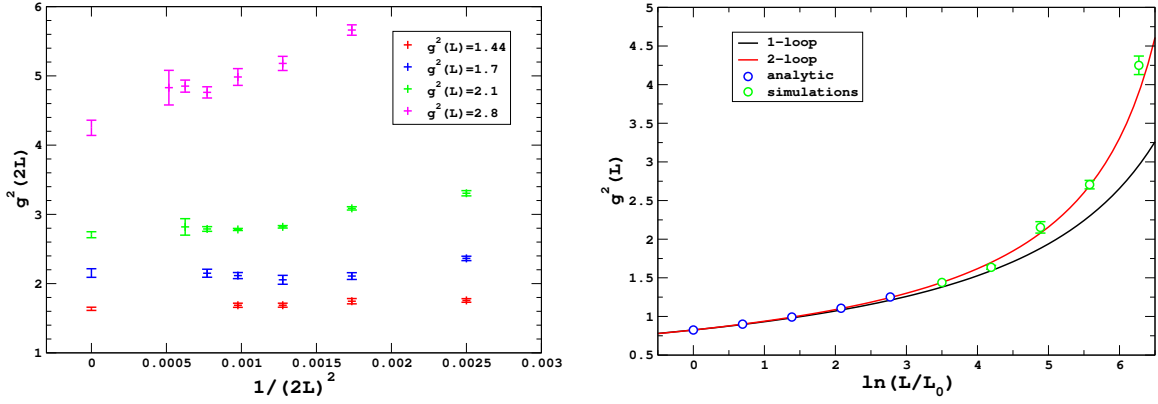

\begin{center}
\begin{tabular}{c c}
\vspace{0.5cm} \hspace{-0.6cm}
\includegraphics[width=0.49\textwidth,angle=0]{gsq2L_all.v5.eps}
\hspace{0.3cm}
\includegraphics[width=0.49\textwidth,angle=0]{running_combined.grace.v2.eps}
\end{tabular}
\end{center}
\vspace{-0.6truecm}
\caption{(Left) The continuum extrapolations of four discrete RG
  steps. (Right) The RG flow $g^2(L_{\rm phys})$, combining analytic
  lattice perturbation theory and the simulation results. The running
  starts at $g^2(L_{{\rm phys},0})=0.825$. There is excellent
  agreement with continuum 2-loop running, at the strongest coupling,
  the simulation results begin to break away from perturbation theory.}
\label{fig2}
\end{figure}

We iterate the procedure, giving four discrete RG steps as
shown in Fig.~\ref{fig2}~(left). At stronger coupling, we need
lattices up to $2L_i=44$ for the continuum extrapolation. The use of
the finite $L$ values of $k$ does not appear to reduce the cutoff
effects. The continuum RG flow is shown in Fig.~\ref{fig2}~(right). At weak
coupling, we use analytic/numeric lattice perturbation theory to calculate the
Wilson loop ratios in finite volumes
\cite{Heller:1984hx,Lepage:1992xa}. The Wilson loops are calculated to
1-loop in the bare coupling in finite volume. The series is reexpanded
in the boosted coupling constant at the relevant scale of the Creutz
ratio. Step-scaling of the finite volume ratios can be used in
exactly the same way as for the simulations results, to determine the
RG flow in the continuum. The analytic RG flow starts at the reference
point $g^2(L_{{\rm phys},0})=0.825$. At weak coupling there is complete agreement
with 2-loop perturbation theory. We connect lattice perturbation theory
to the simulation results by matching the flows at $g^2(L_{\rm phys})=1.44$,
where the simulation RG flow begins. There is continued agreement with
2-loop perturbation theory at even stronger coupling, only at the
strongest coupling do we see deviation from the perturbative flow.

\section{Fundamental fermions}

\begin{figure}
\begin{center}
\vspace{0.5cm}
\includegraphics[width=0.6\textwidth,angle=0]{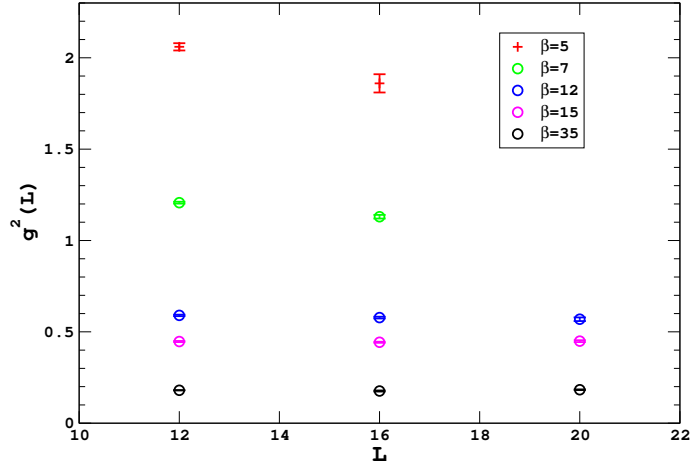}
\end{center}
\caption{\label{fig:nf16}
The renormalized coupling $g^2(L, \beta)$ at fixed bare coupling for
$N_f=16$ fundamental flavors. For $g^2(L, \beta)>0.5$ the renormalized
coupling decreases with increasing $L$, for $g^2(L, \beta)<0.5$ the
renormalized coupling is independent of $L$ within the errors. This is
consistent with the existence of an infrared fixed point.} 
\end{figure}

We next study $SU(3)$ gauge theory with fermions in the
fundamental representation. For $N_f=16$ flavors, 2-loop perturbation
theory predicts the theory is conformal with an infrared fixed point
$g^{*2}\approx 0.5$. Because of the computational expense of
step-scaling with dynamical fermions, in this pilot study we have not
yet extrapolated to the continuum limit. The running coupling
therefore is still contaminated with finite cutoff effects. If the
linear lattice size $L$ is large enough, the trend from the volume
dependence of $g^2(L,\beta)$ should indicate the location of the fixed
point. For $g^2(L,\beta) > g^{*2}$ we expect decrease in the running
coupling as $L$ grows, although the cutoff of the flow cannot be
removed above the fixed point. Below the fixed point with
$g^2(L,\beta) < g^{*2}$ we expect the running coupling to grow as $L$
increases and the continuum limit of the flow could be determined. The
first results are shown in Fig.~\ref{fig:nf16}. We use stout-smeared
\cite{Morningstar:2003gk} staggered fermions \cite{Aoki:2005vt,
  Aoki:2006br} and the RHMC algorithm, 
simulating at quark mass $m_q=0.01$, with some runs at $m_q=0.001$ to
test that the mass dependence is negligible. For the Wilson loop
ratios, we smear the gauge fields and measure the fat-link Wilson
operator. Our experience in the pure gauge theory test is that cutoff
dependence is not reduced using the finite $L$ value of $k$, hence we use
the infinite volume $k$ value to convert the Wilson loop ratios to a
renormalized coupling. The results are
consistent with the above picture. For $g^2(L,\beta)>0.5$, the cutoff
dependent renormalized coupling decreases with $L$. For
$g^2(L,\beta)<0.5$, the renormalized coupling is $L$-independent
within errors. The theory appears conformal but precise determination
of the conformal fixed point requires further studies.

\acknowledgments{We wish to thank Urs Heller for the use of his code
  to calculate Wilson loops in lattice perturbation theory, and Paul
  Mackenzie for related discussions. We are grateful to Sandor Katz
  and Kalman Szabo for helping us in using the Wuppertal RHMC code. In
  some calculations we use the publicly available MILC code, and the
  simulations were performed on computing clusters at Fermilab, under
  the auspices of USQCD and SciDAC, and on the Wuppertal GPU
  cluster. This research is supported by the NSF under grant 0704171,
  by the DOE under grants DOE-FG03-97ER40546, DOE-FG-02-97ER25308, by
  the DFG under grant FO 502/1 and by SFB-TR/55.}


\end{document}